\documentclass[preprint,authoryear,12pt]{elsarticle}

\usepackage{epsfig}

\usepackage{amssymb}

\usepackage[ps2pdf,%
a4paper=true,%
breaklinks=true,%
colorlinks=true,%
pdfauthor={First Author et al.},%
pdftitle={Template for manuscripts in Advances in Space Research}%
]{hyperref}

\journal{Advances in Space Research}

\usepackage{graphicx}
\graphicspath{%
    {converted_graphics/}
    {C:/aa/ddr/stereo/HET/3scplot/}
    {C:/aa/ddr/stereo/HET/NS/}
}
\begin{document}

\begin{frontmatter}



\title{25 MeV solar proton events in Cycle 24 and previous cycles}


\author[ad1,ad2]{Ian G. Richardson\corref{cor}}
\address[ad1]{CRESST and Department of Astronomy, University of Maryland, College Park, 20742}
\cortext[cor]{Corresponding author}
\address[ad2]{Code 661, NASA Goddard Space Flight Center, Greenbelt, MD 20771}
\ead{ian.g.richardson@nasa.gov}


\author[ad2]{Tycho T. von Rosenvinge}
\ead{tycho.t.vonrosenvinge@nasa.gov}

\author[ad3]{Hilary V. Cane}
\address[ad3]{Department of Mathematics and Physics, University of Tasmania, Hobart, Tasmania, Australia}
\ead{hilary.cane@utas.edu.au}

\begin{abstract}
We summarize observations of around a thousand solar energetic particle (SEP) events since 1967 that include $\sim25$~MeV protons, made by various near-Earth spacecraft (IMPs~4, 5, 7, 8, ISEE~3, SOHO), that encompass Solar Cycle 20 to the current cycle (24).  We also discuss recent observations of similar SEP events in Cycle 24 made by the STEREO spacecraft.  The observations show, for example, that the time distribution of $\sim25$~MeV proton events varies from cycle to cycle. In particular, the time evolution of the SEP occurrence rate in Cycle 24 is strongly asymmetric between the northern and southern solar hemispheres, and tracks the sunspot number in each hemisphere, whereas Cycle 23 was more symmetric.  There was also an absence of 25~MeV proton events during the solar minimum preceding Cycle 24 (other minima show occasional, often reasonably intense events).  So far, events comparable to the exceptionally intense events detected in Cycles~22 and 23 have not been observed at Earth in Cycle~24, though Cycle~21 (the largest of the cycles considered here) also apparently lacked such events. We note a correlation between the rates of intense 25~MeV proton events and ``ground level enhancements" (GLEs) observed by neutron monitors, since 1967, and conclude that the number of ``official" GLEs (1) observed to date in Cycle~24 appears to be significantly lower than expected (5 to $7\pm1$) based on the rate of intense 25~MeV proton events in this cycle.

\end{abstract}

\begin{keyword}
Solar energetic particles \sep solar cycle
\end{keyword}

\end{frontmatter}

\parindent=0.5 cm

\section{Introduction}
\begin{figure} 
  \centering

 \includegraphics*[width=11cm,angle=0]{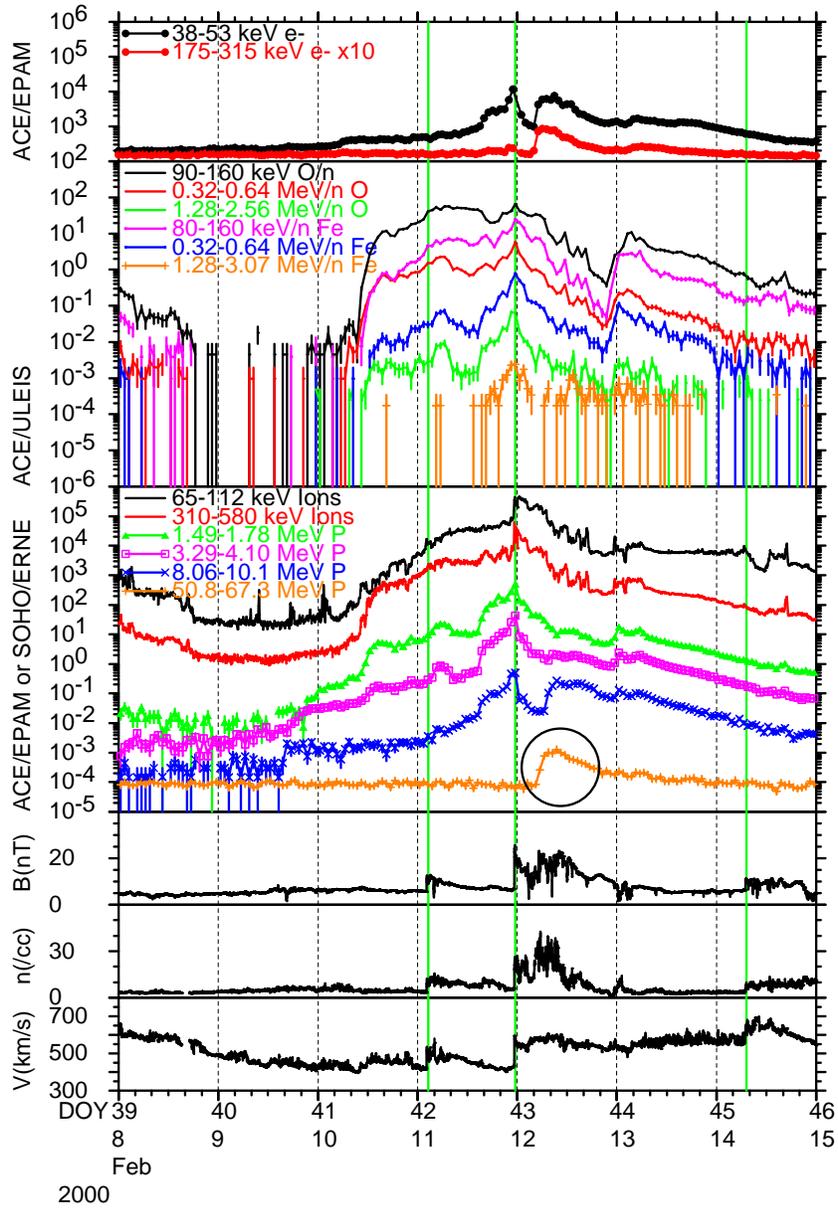}
  \caption{Particle observations from instruments on the ACE and SOHO spacecraft during February 8-14, 2000. Top panel: ACE/EPAM near-relativistic electrons; panel 2: ACE/ULEIS iron and oxygen ions; and panel 3: ACE/EPAM and SOHO/ERNE ions and protons.  Particle intensities are generally in (MeV/n s cm$^2$ sr)$^{-1}$. The bottom panels show the ACE solar wind magnetic field intensity, density and speed, including three shocks (vertical green lines).  A solar particle event onset, most clearly evident in tens of MeV protons, and also in near-relativistic electrons (top panel), is circled. }
  \label{event}
\end{figure}

In this paper, we focus on solar energetic particle (SEP) events that include 25~MeV protons, and summarize some of the properties of such events in Solar Cycle~24 and previous cycles (20--23). We focus on such events for several reasons: First, it is relatively easy to identify SEP event onsets at such energies.  For example, Figure~\ref{event} shows intensity-time profiles at a range of energies for various particle types from instruments on the ACE and SOHO spacecraft located near the Earth during a complex interval in February 2000 when three interplanetary shocks passed the Earth (indicated by green vertical lines; the shocks can be seen in the ACE solar wind magnetic field and plasma observations in the bottom three panels of the figure).   The feature of interest here is the SEP event onset early on February~12 that is clearly evident at proton energies of several tens of MeV (the 51--67 MeV proton intensity from the ERNE instrument on SOHO \citep{t95} is circled), and also in the near-relativistic electron intensity from the EPAM instrument on ACE in the top panel.  It is, however, difficult to identify this onset in the proton and ion intensity-time profiles at energies below a few MeV/n, where particles associated with previous events and local shocks are dominant.  Having identified this event, the dispersive onset may then be traced to lower energies, in particular in the heavy ions observed by the ULEIS instrument on ACE in panel 2.  The tens of MeV proton and near-relativistic electron onset times link the SEP event to an M1.7 X-ray flare with peak intensity at 0410~UT, located at  N26$^\circ$W23$^\circ$ and associated with a 1107 km~s$^{-1}$ full halo coronal mass ejection (CME) in the CDAW CME catalog (\url{http://cdaw.gsfc.nasa.gov/CME\_list/}) \citep{c10}.  As Figure~\ref{event} exemplifies, proton observations below a few tens of MeV, including the GOES $>10$ MeV proton intensity used (with a threshold of 1 (cm$^2$ sr s)$^{-1}$) to compile the NOAA SEP event list (\url{ftp://ftp.swpc.noaa.gov/pub/ indices/SPE.txt}), may be dominated by particles associated with local interplanetary shocks which can obscure new event onsets that are evident at higher energies (see also \cite{c07}).

A second reason to focus on 25~MeV proton events is that we have routinely identified such events since 1967 in observations from the Goddard Space Flight Center (GSFC) and other instruments on various spacecraft, and have used them in several studies. These include: \cite{c88}, where observations from 235 events in 1967--1985 were used to demonstrate clearly the contribution of interplanetary shocks to the distribution of SEPs in longitude; \cite{c10}, which discussed 280 events in 1997--2006 during Cycle 23; and \cite{r14}, which summarized the properties of 209 25 MeV proton events observed by the STEREO A or B spacecraft and/or near the Earth in 2006--2013. 

Such events are also generally associated with relatively energetic solar events.  For example  (\cite{c10}; \cite{r14}), essentially all 25~MeV proton events are accompanied by CMEs, X-ray flares (if originating on the front side of the Sun), type III radio emissions, and less frequently, by type II radio bursts ({\it e.g.}, 53\% of the events of \cite{r14} were accompanied by type II bursts identified in WIND/WAVES and/or STEREO/SWAVES observations (\url{http://www-lep.gsfc.nasa.gov/waves/})).

In this study, we use this extensive catalog of SEP events including 25~MeV protons to summarize a few properties of such events in Cycle~24 and in previous cycles.  An important point to note is that we do not require the proton intensity to exceed an arbitrary threshold for a particle increase to be considered as an ``event", such as is required for inclusion in the NOAA SEP event list referred to above.  Rather, we include all proton enhancements that are detectable above low instrumental backgrounds.  Hence, we consider a wider dynamic range in intensity than studies that focus just on larger events, such as those in the NOAA list.  We also reject intensity enhancements that are likely to be modulations of ongoing events rather than true solar events based on, for example, a lack of velocity dispersion, no obvious solar signatures, and the presence of local solar wind structures.

\begin{figure}
 \centering
 \includegraphics*[width=13cm,angle=0]{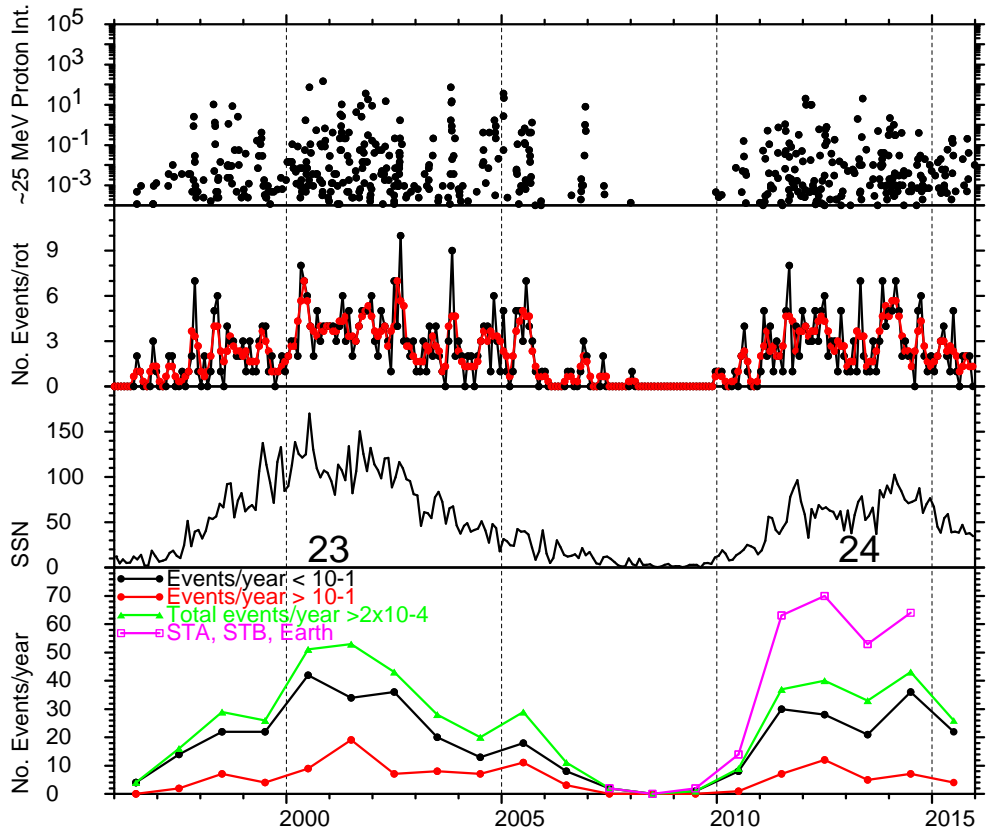}
  \caption{Top panel: Peak intensities in (MeV/n s cm$^2$ sr)$^{-1}$ of 25~MeV proton events observed at Earth during Solar Cycles~23 and 24, to 2015; Panel 2: the number of proton events at Earth/Carrington rotation and the 3-rotation running mean; Panel 3: The sunspot number (SSN; original version) from WDC-SILSO; Bottom panel: Total yearly number of proton events detectable at 25~MeV at Earth (green graph), and the number of events above (red graph) and below (black graph) an intensity of $10^{-1}$~(MeV s cm$^2$ sr)$^{-1}$. The number of individual events detected at the STEREO A/B spacecraft and/or at Earth is indicated for 2007--2014 (purple graph). }
  \label{earthrate}
\end{figure}  

\section{Solar Cycles~23 and 24}
For Solar Cycle 24, we use an updated version \citep{r16} of the catalog of 25~MeV proton events compiled by \cite{r14} using observations from the High Energy Telescopes \citep{vr08} on the STEREO~A (``Ahead") and B (``Behind") spacecraft \citep{k08} and from near-Earth spacecraft, since launch of the STEREO spacecraft on October~26, 2006.  STEREO A and B moved ahead or behind the Earth in its orbit, respectively, advancing at $\sim22^\circ$/year, and passed each other on the far side of the Sun in March 2015.  

For direct comparison with the events observed by near-Earth spacecraft in Cycle~23  \citep{c10}, we will generally focus here on the Cycle~24 events that were observed near the Earth.  Figure~\ref{earthrate} is an updated version of Figure~10 of \citet{r14} which shows in the top panel the peak 25 MeV proton intensities for events observed at Earth in 1996--2015.  The second panel shows the number of 25 MeV proton events/Carrington rotation and the three-rotation running averages.  These rates tend to track the monthly-averaged sunspot number, from the World Data Center for Sunspot Index and Long-term Solar Observations (WDC-SILSO) at the Royal Observatory of Belgium, Brussels, shown in the third panel.  (A revised sunspot series has recently been released by SILSO \citep{cl14}, but here we use the original version which is essentially a factor of $1/0.6$ larger than the revised series for the period used in this paper.  Points after May 2015, the end of the original series, are from the revised series and scaled by this factor.)  The interval shown includes Cycle~23 and the rising, peak and early declining phases of Cycle~24.  The bottom panel shows the number of events/year detected above an instrumental threshold of $\sim2\times10^{-4}$~(MeV~s~cm$^2$~sr)$^{-1}$ (green graph), and the numbers of these events with intensities above (red graph) and below (black graph) a threshold of $10^{-1}$~(MeV~s~cm$^2$~sr)$^{-1}$.  The total number of individual events observed by the STEREO spacecraft and/or at the Earth, from \cite{r14} and updated with events in 2014, is also shown (purple graph).  By individual events, we mean that an event observed at more than one location is only counted once.  No rate is shown for 2015 since contact was lost with STEREO B on October 1, 2014, and data were only received intermittently from STEREO A, located behind the Sun as viewed from Earth \citep{r16}.  

The yearly event rates in Figure~\ref{earthrate} reflect the smaller Cycle~24 compared to Cycle~23. In particular, 2001 had the largest number of 25 MeV proton events (53) detected at Earth in Cycle~23, whereas the largest number so far in Cycle~24 is 43~events during 2014.  Considering the first seven years of each cycle, starting from the cycle onsets (based on minimum smoothed sunspot numbers) in May, 1996 and December, 2008, 231 events were detected at Earth in Cycle~23 compared with 188 in Cycle~24, while 48 and 36 events with peak 25 MeV proton intensities $\ge10^{-1}$~(MeV s cm$^2$ sr)$^{-1}$ were observed in the respective cycles.  Thus, in both cases, the 25~MeV proton event rate in Cycle~24 was around 80\% of that in Cycle 23. If we consider a higher intensity threshold of 1~(MeV s cm$^2$ sr)$^{-1}$, there were 21 events during the first seven years of Cycle 23 (note that this interval ends before the intense ``Halloween" events in October--November, 2003 ({\it e.g.}, \cite{m05} and references therein) compared with 9 in Cycle 24, or 43\% of the Cycle 23 rate.   

Considering events in 2009--2015, 189 25~MeV proton events were observed at Earth.  However, an additional 77 individual events were detected only at one or both of the STEREO spacecraft \citep{r14}.  Thus, around 29\% (77/266) of these 25~MeV proton events were not detected at the Earth.  Unlike the event rate at Earth, which peaked in 2014, the combined near-Earth and STEREO rate peaked in 2012.  However, it is likely that the 2014 rate is depressed by the loss of STEREO B data and intermittent STEREO A observations when on the far side of the Sun ({\it c.f.}, \cite{r16}).

\begin{figure}
 \centering
\includegraphics*[width=12cm,angle=0]{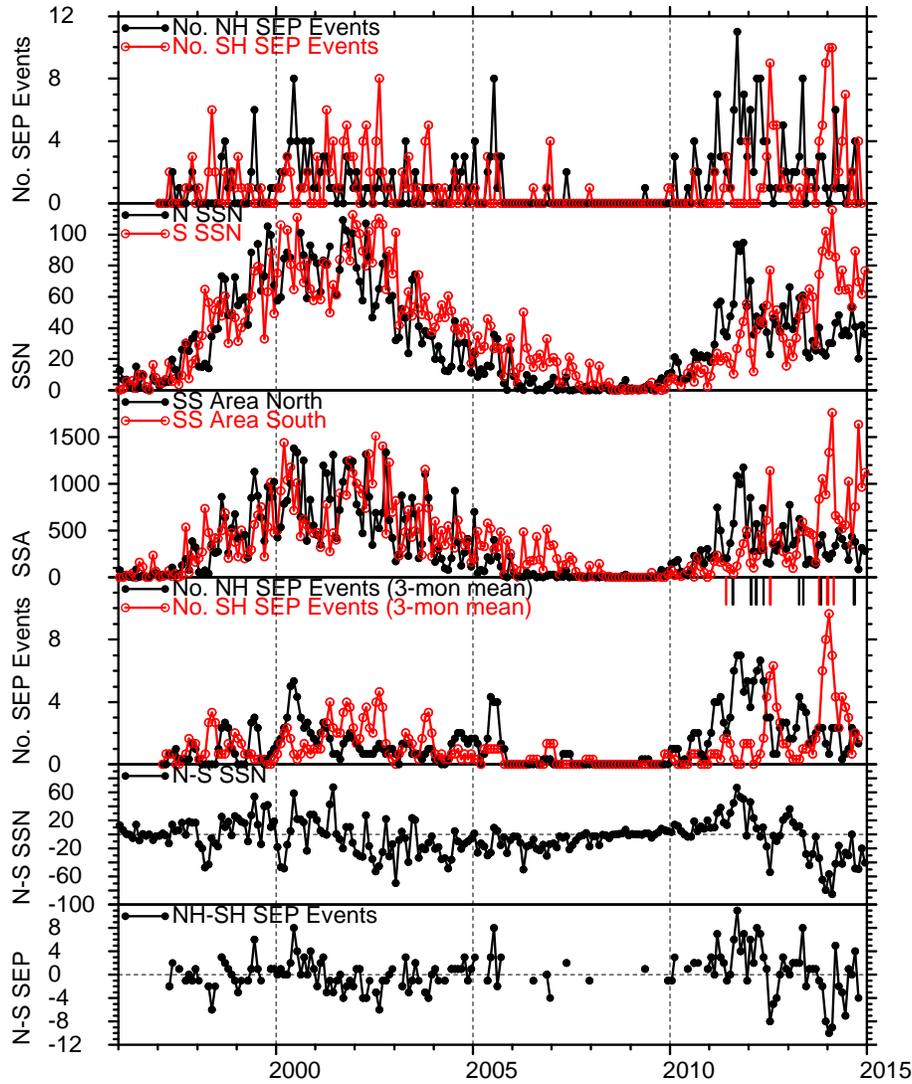}
  \caption{The top panel shows the number of 25~MeV proton events/month detected at Earth in 1997--2006 and, from 2007, also including events detected at the STEREO spacecraft, related to solar events in the northern (black) or southern (red) solar hemispheres. The monthly hemispheric sunspot numbers and Greenwich/USAF/NOAA sunspot areas (in $\mu$-hem) are shown in the second and third panels, respectively.  Panel 4 shows 3-month running means of the SEP event occurrence rates for each hemisphere. Tick marks indicate the time and hemisphere of SEP events observed by AMS--02 \citep{bin15}.  The strong hemispheric asymmetry in activity (SEP events and sunspots) in Cycle 24 contrasts with the more symmetrical situation in Cycle 23. The bottom two panels show the differences in the northern and southern sunspot numbers or SEP rates, indicating a change in Cycle 24 from a predominantly northern bias to predominantly southern around May 2013.  Cycle 23 shows a more modest bias, also turning (in 2001) from northern to southern.}
  \label{ns}
\end{figure} 

In Figure~\ref{earthrate}, Cycle 24 shows two peaks in the SEP rate, in 2011--2012, and 2014, associated with the two sunspot peaks in this cycle.  These are separated by a temporary decrease in the occurrence of 25~MeV proton events that is probably a manifestation of the ``Gnevyshev gap" (GG), a temporary decrease in energetic solar activity, including SEP events, often found near solar maximum ({\it e.g.}, \cite{g67}, \cite{g77}, \citet{s03}; \citet{ng10}). This feature is examined in more detail in Figure~\ref{ns} where we show the 25~MeV proton event rate/month for Cycle~24 (and also Cycle 23) in the top panel, separating the events by whether they originated in the northern (black) or southern (red) solar hemisphere based on the location of the related solar activity (see \citet{r16} for further details). To increase the SEP event statistics in Cycle 24, we include here events detected by the STEREO spacecraft as well as those observed at the Earth.  For sources on the far side of the Sun from Earth, the hemisphere is inferred from STEREO imaging data.  

For comparison with the SEP rates, the second and third panels show the monthly-averaged hemispheric sunspot numbers and the Greenwich/USAF/NOAA sunspot areas (in millionths of the visible hemisphere, $\mu$-hem) compiled by David Hathaway 
(\url{http:// solarscience.msfc.nasa.gov/greenwch.shtml}), respectively.  It is evident that the number of proton events in each hemisphere in Cycle 24 closely follows the sunspot number/area in the respective hemisphere, including the first sunspot peak, dominated by northern sun spots and particle events, and the second peak, dominated by southern sun spots and particle events.  Thus, the Gnevyshev gap in Cycle~24, evident in the reduced SEP event rates in the interval between the northern and southern sunspot peaks, is a period of transition between the declining northern activity and the increasing southern activity.  Other intervals when sun spots and SEP events from a particular hemisphere are dominant are evident, including a brief interval of enhanced southern activity in 2012 that follows the northern hemisphere activity peak and included the largest SEP event observed (at STEREO A) so far in this cycle \citep{ru13}.  In addition, there are quasi-periodic $\sim6$--7 month (i.e., data point) variations in the SEP rates in each hemisphere that are also evident in the respective sunspot number and areas and are present during much of the rise and peak phases of Cycle 24, as discussed in more detail by \cite{r16}.  

The fourth panel of Figure~\ref{ns} shows 3-month running means of the SEP rates in each hemisphere.  Tick marks indicate the time and hemisphere of the 20 solar particle events observed by AMS-02 \citep{ag15} at rigidities near 1~GV and above during the first three and a half years after commencing operation on May~19, 2011 \citep{bin15}. These follow a similar pattern in terms of their temporal and hemispheric distributions as the lower energy proton events.  The bottom two panels show the differences between the northern and southern sunspot areas or the number of SEP events in each hemisphere.  Both show a predominant bias towards the north during the leading part of Cycle~24, turning to a predominantly southern bias from around May~2013. 
 
Figure~\ref{ns} shows a clear contrast between the strong north-south asymmetry in Cycle~24 and the more symmetric situation in Cycle~23.  Although Cycle~23 also had two sunspot peaks (Figure~\ref{earthrate}), with a GG indicated by the reduction in sunspot number and area in late 2000--early 2001, each peak involved both hemispheres.  There is only a small bias (panel 5) towards the northern hemisphere during the rise and early peak in the cycle, and a southern bias in the second peak and declining phase through to the start of Cycle~24. Thus, the north-south bias evolved similarly with time in both these cycles even though the global solar magnetic field reversal was oppositely-directed in each cycle.  This bias is also evident in the SEP events (bottom panel).  Because only events detected at Earth are considered for Cycle~23, the statistics are relatively smaller than for Cycle~24, for which STEREO events are also included in Figure~\ref{ns}, even though event rates in Cycle~23 were actually higher as discussed above. However, with the available events, the asymmetry between hemispheres is less pronounced in Cycle~23 than in Cycle~24.  Fluctuations in the SEP rate and sunspot numbers/areas are also evident in Cycle~23, and periodic variations in the occurrence of SEP events and in other parameters in Cycle~23 have been reported (e.g., \cite{d01},  \cite{rc05}).  

\begin{figure}  
  \centering
  \includegraphics*[width=13cm,angle=0]{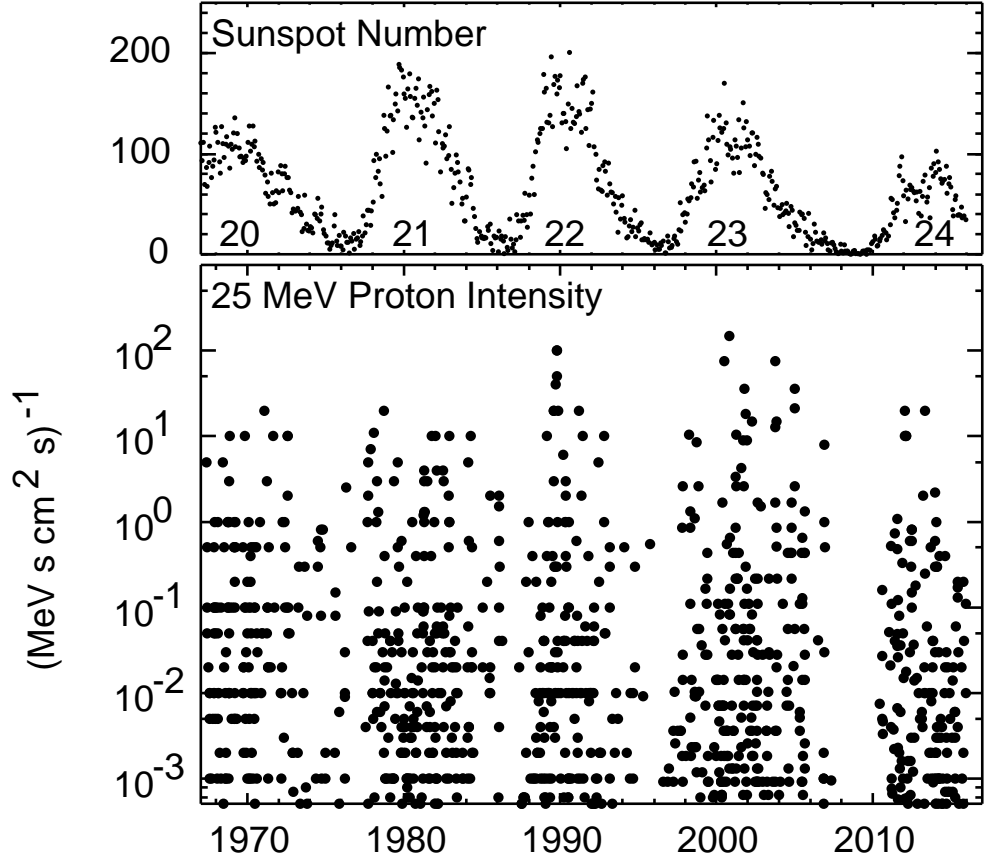}
  \caption{Peak intensities of 1008 25~MeV proton events detected at Earth in 1967--2015 with intensities $\ge5\times10^{-4}$ (MeV s cm$^2$ sr)$^{-1}$.  The top panel shows the monthly sunspot number.  Intensities prior to 1972 are subject to saturation of the early IMP instruments, and are estimated by eye from plots in \cite{v74}.  Note the absence of 25~MeV proton events in the solar minimum between Cycles~23 and 24, and also the absence of intense events comparable to those in Cycles~22 and 23 in both Cycles~21 and 24. }
  \label{25issn}
\end{figure}

\section{25~MeV proton events since 1967}

We now broaden the discussion to include observations of near-Earth 25~MeV proton events since 1967, an interval spanning Solar Cycles~20 to 24.  As discussed above, our starting points were the 235 events from 1967 to 1985 used by \cite{c88} and the recent event lists of \cite{c10} (covering 1997--2006) and \cite{r14} (from 2006, updated to near present).  We have also perused observations from the GSFC instruments on the IMP~4 (data for 1967--1969), 5 (1969--1972), 7 (1972--1978), 8 (1973--2006), and ISEE-3 (1978--1983) spacecraft near the Earth to restore all the detectable 25~MeV proton events that were not included in the \cite{c88} study.  These include weak events (their study required a threshold of $3\times 10^{-3}$ (MeV s cm$^2$ sr)$^{-1}$ at 9--23~MeV), events without well-identified solar sources (possibly on the far-side), and those with data gaps, that were not selected by \cite{c88} but should be included in a survey of events. We have also filled a gap between these lists in 1986--1996, including Cycle~22. This catalog is a work in progress for future publication, and the limited aim of this paper is to summarize a few results that may be obtained with this large data set.  In particular, identification of the solar sources of the added/restored events is not yet complete and there are limitations, such as data coverage and instrument capabilities ({\it e.g.}, saturation of the IMP~4 and 5 instruments in large events) that may be corrected by examining data from other instruments/spacecraft. In addition, 25~MeV proton event intensities identified using IMP~4 and 5 data are currently estimated by eye from intensities at adjacent energies (6--19, 19--80 MeV) in the plots of \cite{v74}.  Note that IMP~8 observations extended from October, 1973 to 2006, so many events in Cycles 20--23 were observed by the same instrumentation (the Goddard Medium Energy instrument (GME); \url{http://spdf.gsfc.nasa.gov/imp8\_GME/GME\_home.html}).

Combining these sources gives a current total of 1147 25~MeV proton events detected near the Earth in 1967--2015. Figure~\ref{25issn} shows the 1008 events with peak intensities $\ge5\times10^{-4}$ (MeV s cm$^2$ sr)$^{-1}$ (a slightly higher threshold is used here than for the Cycle~23 and 24 events discussed above to allow for the higher background in the early IMP instruments).  The top panel shows the monthly sunspot number, indicating that this interval extends from the rising phase of Cycle~20.  The occurrence of the proton events clearly follows the solar activity cycle, as will be discussed further below.  Comparing Cycle~24 with previous cycles, the most intense events observed at Earth so far are around an order of magnitude smaller than those detected in both Cycles~23 and 22.  However, it is notable that such extreme events were also absent (at least at Earth) in Cycle 21, the largest cycle of the space era; the largest Cycle~21 events are comparable to those found in Cycle 24.  Hence, the absence of the largest proton events at Earth in Cycle~24 compared to Cycle~23 may not simply be a reflection of the weaker Cycle~24 but also that Cycle~23, despite being weaker than the previous two cycles, nevertheless produced exceptionally intense particle events.  Note also that, as already mentioned, IMP~8 made observations in Cycles~21, 22 and 23, so the lack of large events in Cycle 21 is not due to an instrumental limitation.   Considering Cycle~20, which was weaker than the subsequent two cycles, the intensities (from IMPs~4 and 5) are limited by instrument saturation.  In particular, the exceptionally intense August 4--5, 1972 event (e.g., \cite{pd74}; \cite{le76}) is not evident in Figure~\ref{25issn}.

\begin{figure} 
  \centering
   \includegraphics*[width=13cm,angle=0]{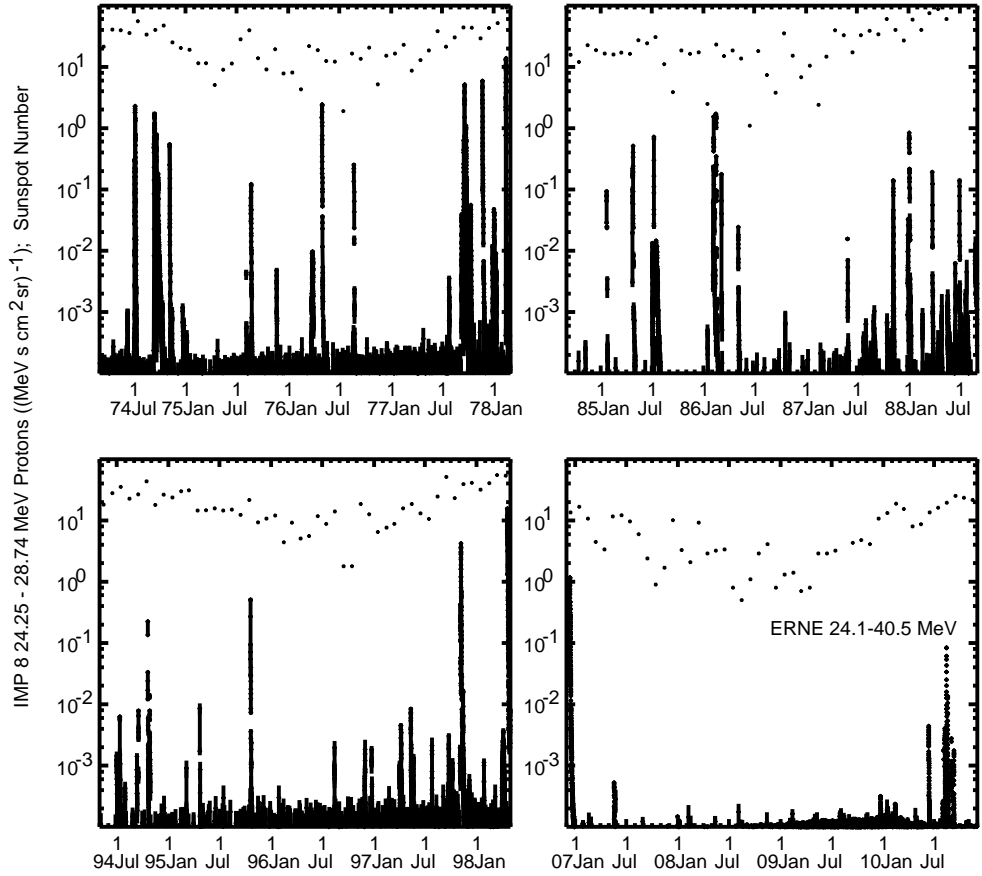}
  \caption{The intensity of 24.25--28.74~MeV protons observed by GME on IMP 8 during $\pm2$~years around the smoothed sunspot minima between Solar Cycles~20 and 21 (top left), 21 and 22 (top right) and 22 and 23 (bottom left), and of 24.1--40.5~MeV protons detected by ERNE on SOHO between Cycles~23 and 24 (bottom right).  The monthly sunspot number is shown by dots.  The lack of any substantial $\sim25$~MeV proton events during the minimum between Cycles~23 and 24 contrasts with previous minima when relatively intense events were occasionally observed. }
  \label{min}
\end{figure}

Another feature evident in Figure~\ref{25issn} is the absence of any substantial 25~MeV proton events in the solar minimum between Cycles~23 and 24, whereas such events were detected during other solar minima in this figure.  Figure~\ref{min} shows $\sim25$~MeV proton intensities observed during $\pm2$~years around the time of smoothed sunspot minimum in each of the minima in Figure~\ref{25issn} and illustrates that this lack of events is unusual.  More usually, solar particle events (including some ``ground level enhancements" (GLEs) that are sufficiently energetic that solar particles are detected by neutron monitors, {\it e.g.}, \cite{a11}, and references therein) are observed at least occasionally during solar minima ({\it e.g.},  \cite{ss89}).  The monthly (un-smoothed) sunspot number is also shown (by dots)  in Figure~\ref{min}, the logarithmic scale accentuating the differences in the lowest sunspot numbers during each minimum.  The absence of such proton events in the minimum between Cycles~23 and 24 was probably associated with the unusually low sunspot number ($\ll 10$) compared with $\sim10$ or more in previous cycles when active regions occasionally formed and produced a 25~MeV proton event. For example, particle events during the minimum between Cycles~20 and 21 have been discussed by \cite{k93}.  Thus, the absence of significant energetic particle events is another unusual characteristic of the solar minimum between Cycles~23 and 24 compared with previous minima in the space era.

\begin{figure} 
  \centering
  \includegraphics*[width=12cm,angle=0]{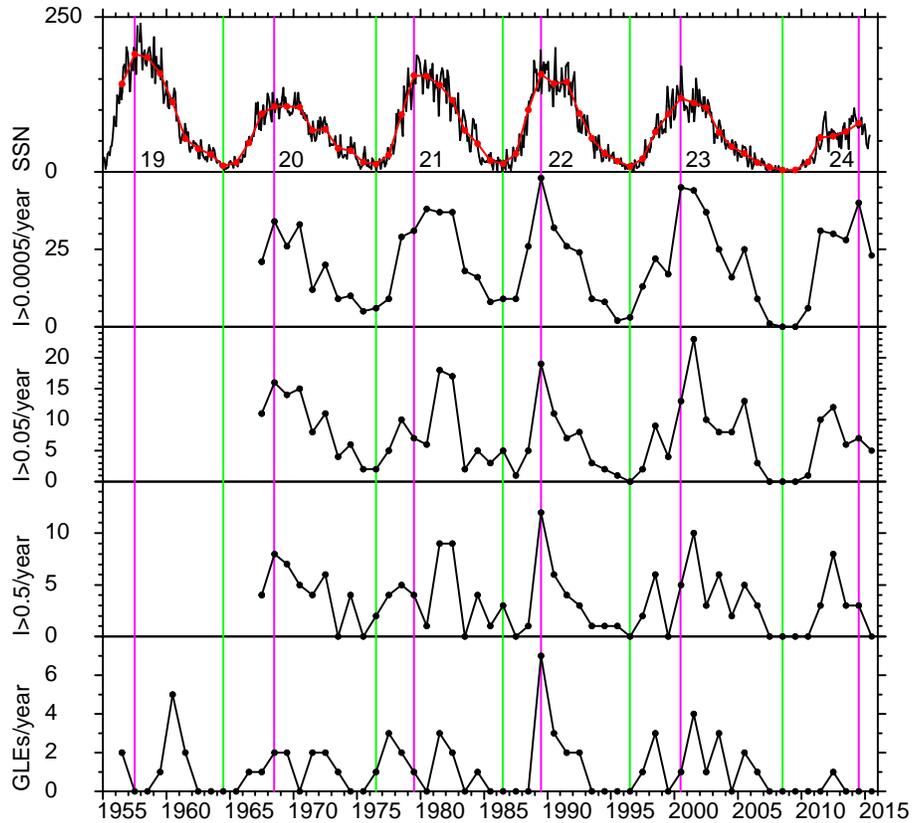}
  \caption{The top panel shows the monthly and annual sunspot numbers for 1955--2015 encompassing Cycles~19 to 24. The next three panels show the number of 25~MeV proton events identified/year with $\sim25$~MeV proton intensities $\ge0.0005$, $\ge0.05$ and $\ge0.5$ (MeV s cm$^2$ sr)$^{-1}$.  The bottom panel shows the number/year of ``official" ground level enhancements observed by neutron monitors (\url{https://gle.oulu.fi/}).  }
  \label{longrate}
\end{figure}

Figure~\ref{longrate} summarizes the number of proton events identified each year with intensities at 25~MeV in three intensity ranges,  $\ge0.0005$, $\ge0.05$ and $\ge0.5$ (MeV s cm$^2$ sr)$^{-1}$, shown in panels 2--4 respectively; panel~1 shows the monthly and annual-averaged sunspot number.  No corrections have been made to the proton event rates to allow for data coverage, which however, mostly affects the pre--IMP 7/8, data.  The bottom panel shows the number of ground level enhancements in each year from the official GLE list, currently maintained at the University of Oulu (\url{https://gle.oulu.fi/}).  Since this starts in 1956, the sunspot number in the top panel is extended back to 1955 to include Cycle~19.  Green and purple lines indicate years of minimum and maximum yearly sunspot number, respectively.  

The close relationship between the solar cycle and the number of detected 25~MeV proton events is clearly evident in the top two panels, and in most cycles (21 is the exception), the maximum number of proton events occurs in the year of sunspot maximum.  For more intense events, while there is also a solar cycle variation, the occurrence rate is less well correlated with the sunspot number.  In particular, Cycle~21 shows a significant fall in the occurrence of large proton events in 1980, the year after sunspot maximum, related to the GG in this cycle \citep{fs97}. (Other parameters such as geomagnetic activity, the solar mean field, and the interplanetary magnetic field, were also depressed at this time, e.g., \cite{rc12}).  This feature is also evident in the GLE rate in Cycle~21; Cycle~19 likewise shows a decrease in the GLE rate at cycle maximum, again indicative of a GG (see also \cite{n91}). On the other hand, Cycles~22 and 23 do not show similar decreases in the occurrence of large particle events at solar maximum, but rather increases.  (Note that the most prominent rate decrease in Cycle~23, in 1999, occurred during the rise phase and is not related to the GG; see also the discussion of Figure~\ref{ns} above.) 

\begin{figure}
  \centering
    \includegraphics*[width=13cm,angle=0]{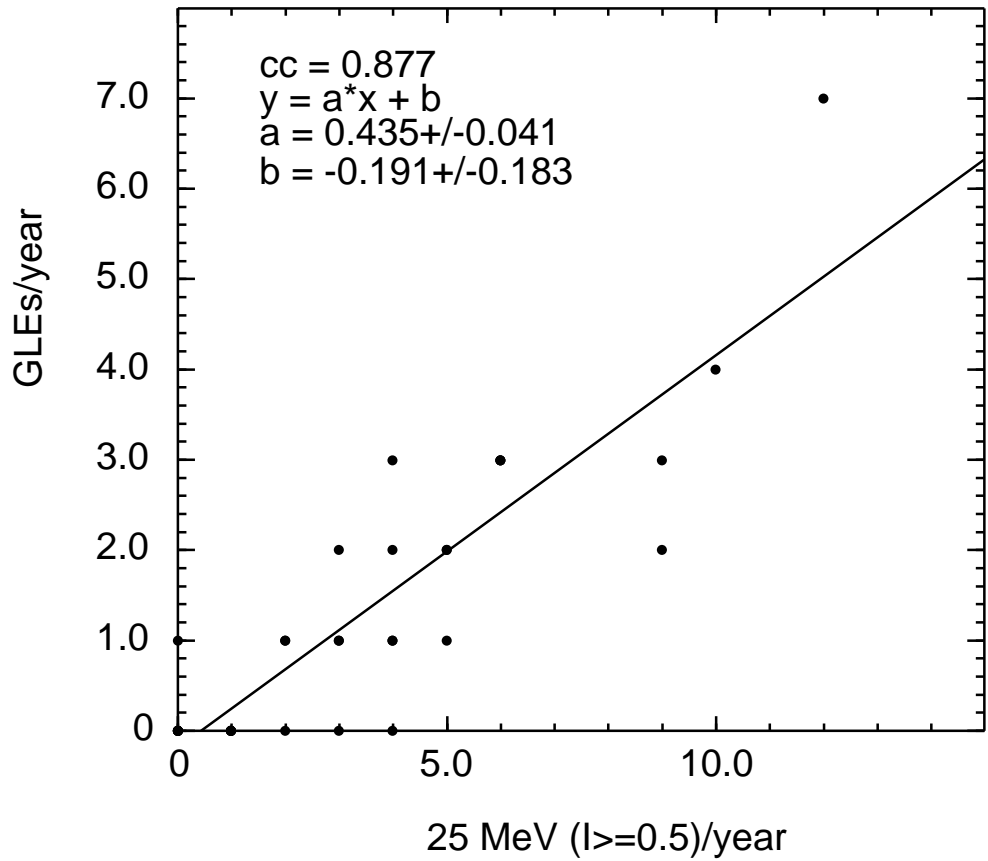}
  \caption{The number of GLEs/year in 1973--2007 plotted vs. the number of 25~MeV proton events with intensities $\ge0.5$ (MeV s cm$^2$ sr)$^{-1}$ identified in the same year.  }
  \label{gle05}
\end{figure}

It is also evident in Figure~\ref{longrate} that the GLE rate generally closely tracks the rate of the largest 25~MeV proton events ($I\ge0.5$ (MeV s cm$^2$ sr)$^{-1}$) until Cycle 24, where a deficiency of GLEs relative to the 25~MeV proton event rate is evident.  Only one official GLE (May~17, 2012) is recognized in Cycle~24 at the time of writing, though \cite{t14} have argued that a second ``GLE" occurred on January~6, 2014 but did not meet all the usual requirements for official recognition, in particular, detection by more than one neutron monitor.  To quantify the lack of GLEs in Cycle~24 using the observations in Figure~\ref{longrate}, Figure~\ref{gle05} shows the number of GLEs/year plotted against the number of $I\ge0.5$ (MeV s cm$^2$ sr)$^{-1}$ 25~MeV proton events/year from 1973 ({\it i.e.}, after IMP~7 launch, to avoid saturation and data gaps in data from the early IMPs) to 2007, preceding Cycle~24.  The linear least-squares fit in the figure suggests that the GLE rate is around $44\pm 4$\% of the $I\ge0.5$ (MeV s cm$^2$ sr)$^{-1}$ 25~MeV proton event rate at Earth.  Since 17 such proton events were observed at Earth in Cycle~24 to 2015, this result suggests that around $7\pm1$ GLEs might have been expected in Cycle~24 by the end of 2015, rather than the one (or maybe two) actually observed, indicating a deficiency of extremely energetic ($\sim$GeV) SEP events relative to intense $\sim25$~MeV proton events observed at Earth in Cycle~24.  (A similar point is made by \cite{b15} from comparing the rates of GLEs and $>10$ and $>100$~MeV proton events in Cycles 21--24.)  

The linear least-squares fit in Figure~\ref{gle05} does not include any uncertainties in the annual numbers of GLEs and proton events.  If we make the simplest assumption that the uncertainties are equal to the square root of the yearly numbers of GLEs or SEP events in Figure~\ref{gle05}, this gives a weighted linear fit ($a=0.352\pm0.044$, $b=-0.107\pm 0.143$, correlation coefficient ($cc$) $= 0.811$) that suggests an expected $\sim6\pm1$ GLEs in Cycle~24 to the end of 2015.  If the highest (1989) point with 7 GLEs is removed and the fits recalculated, the expected number of Cycle~24 GLEs is $6\pm1$ ($cc=0.845$) without including errors in the event numbers and around $5\pm1$ with errors included ($cc=0.795$).  These estimates are still significantly higher than the one, or possibly two, GLEs actually observed in Cycle 24 to 2015, suggesting that the conclusion that the GLE rate is Cycle 24 is lower than expected based on the 25~MeV proton event rate is reasonably robust.  We note however, that the statistics of rare extreme events requires careful consideration, as discussed by \cite{r12}.

We have also considered whether the Earth might have been in an unfavorable position to observe GLEs in Cycle 24,  Although this cannot be firmly established since GLEs are by definition only observed at Earth, the fact that STEREO~A and B observed similar numbers of $I\ge0.5$ (MeV s cm$^2$ sr)$^{-1}$ 25~MeV proton events (21 and 17, respectively to the end of 2014) as the Earth (17) suggests that the Earth was not in an especially unfavorable location to observe intense particle events.  On the other hand, large SEP events including GLEs (as is evident from perusal of the GLE list) frequently occur in ``episodes" that are often associated with a major active region ({\it e.g.}, \cite{ss90}).  Thus, it is also possible that the low GLE rate is due to the Earth being unfavorably located to detect such episodes of GLE production in Cycle 24.       
          
\begin{figure}
  \centering
    \includegraphics*[width=13cm,angle=0]{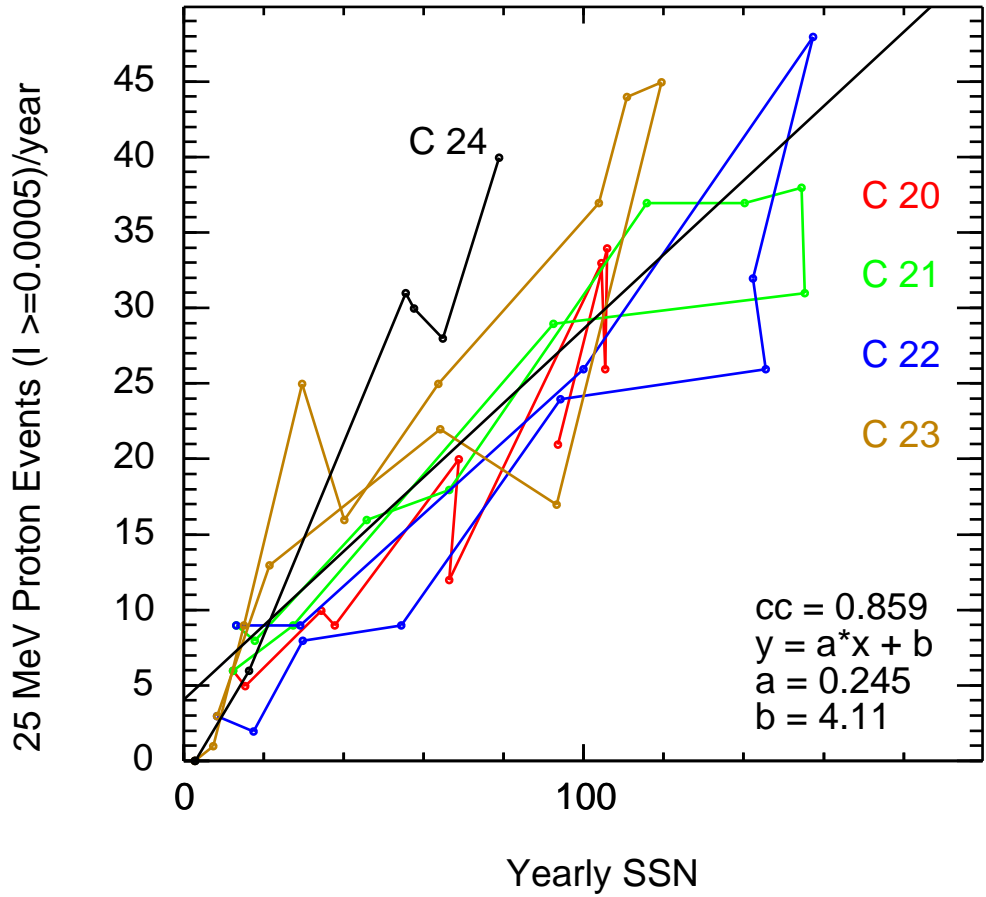}
  \caption{The annual number of 25~MeV proton events with intensities $\ge0.0005$~(MeV~s~cm$^2$~sr)$^{-1}$ identified at Earth plotted against the average sunspot number in the same year, for 1967--2014 with each solar cycle indicated.  Relative to the sunspot number, the 25~MeV proton event rate in Cycle 24 has trended above the rates for Cycles 20 to 23.  The linear fit includes all the data shown in the figure.}
  \label{detssn}
\end{figure}

Figure~\ref{detssn} shows the good correlation ($cc=0.859$) between the total number of 25~MeV proton events detected at Earth/year (with intensities $\ge5\times10^{-4}$ (MeV s cm$^2$ sr)$^{-1}$) and the mean sunspot number in the same year, from 1967 to 2014 ({\it cf.}, the top two panels in Figure~\ref{longrate}).  The line color indicates the solar cycle number.  This plot suggests that, at least relative to the sunspot number, the number of 25 MeV proton events at Earth in Cycle~24 was actually higher than typically observed in previous cycles by a factor of around two, a conclusion also reached by \cite{b15} by comparing the accumulated $>10$ and $>100$~MeV proton events and sunspot numbers during Cycle~24 compared to Cycles~21--23.  Thus, Cycle~24 was not particularly lacking in energetic proton events relative to previous cycles except for the highest energy (GLE) events and also the intense events that were detected in Cycles~22 and 23, but not in Cycle~21.       

\begin{figure} [t] 
  \centering
   \includegraphics*[width=11cm,angle=0]{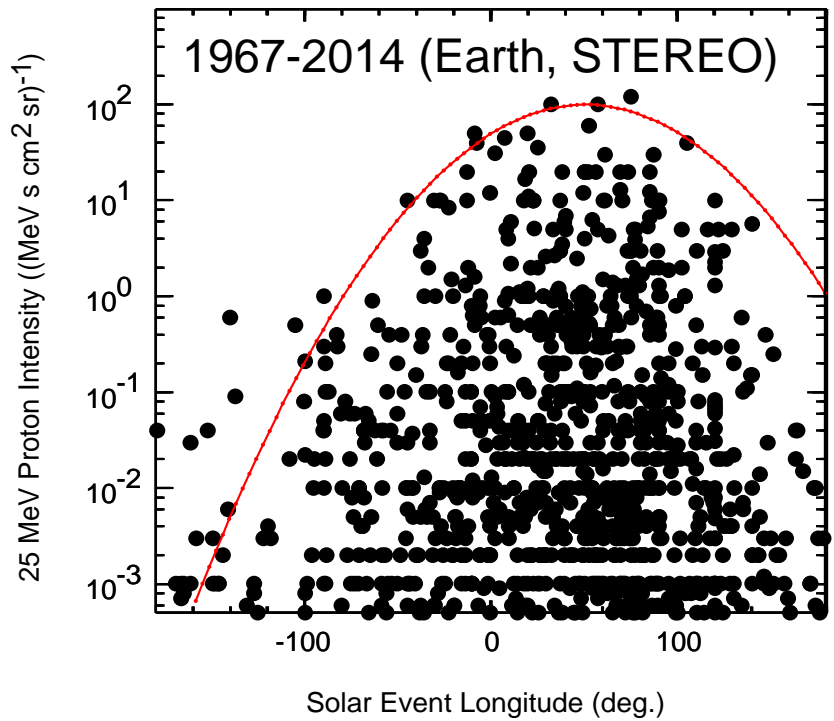}
  \caption{Observed peak 25~MeV proton intensity plotted against the longitude of the related solar event relative to the observer (positive = west), for events in 1967--2014 including observations at Earth and Cycle 24 observations from the STEREO spacecraft.  A total of 951 observations are included in the figure.  The Gaussian shown for comparison has the typical width found for the ``three-spacecraft" (STEREO A, B and near-Earth) events reported by \cite{r14}, is centered at the 450~km~s$^{-1}$ solar wind connection longitude, and the height is aligned with the largest events observed.  }
  \label{intlon}
\end{figure}

Figure~\ref{intlon} summarizes the observed 25~MeV proton peak intensities plotted against the longitude of the related solar event relative to the observing spacecraft (positive = western longitudes) for 951 events in 1967--2014 for which the event location is known.  In addition to proton events detected at the Earth, we have included events in Cycle~24 observed by STEREO~A or B in the figure. As discussed by \cite{r14}, STEREO observations have demonstrated conclusively that 25~MeV proton events can originate anywhere on the Sun relative to the observing spacecraft, as is evident in Figure~\ref{intlon}.  For proton events in previous cycles originating on the far side of the Sun, the longitude of the related solar event usually cannot be estimated except in cases when the solar event was most likely in an active region that had recently passed over the west limb or rotated over the east limb following the proton event.  In addition, for the Cycle 23 proton events discussed by \cite{c10}, far side events were identified using CME and solar imaging (showing an absence of front side activity), and high frequency occultations of type III radio bursts due to the limb of the Sun. Default locations ({\it e.g.}, W120$^\circ$) were assigned by \cite{c10} unless the location could otherwise be estimated from active region rotations over the limb, and these are retained in Figure~\ref{intlon}.  

Figure~\ref{intlon} clearly shows the western biases in the proton intensities and the solar event distribution introduced by the Archimedean-spiral interplanetary magnetic field, which best connects to solar events on the western hemisphere relative to the observer.  The proton intensity fall-off with increasing longitudinal separation from well-connected western sources noted in other studies ({\it e.g.}, \cite{k93}; \cite{lar06}, \cite{lar13}; \cite{r14}) is also evident. In particular, the red curve is a Gaussian centered at the 450~km~s$^{-1}$ solar wind connection longitude with a peak that is consistent with the intensities of the largest observed events.  The width is given by $\sigma=43^\circ$ (full width at half maximum $= 2.355\sigma$), the average value found for a sample of  ``three-spacecraft" 25~MeV proton events observed by both STEREO spacecraft and at the Earth by \cite{r14}; see also \cite{lar13}. This curve is not a fit to the data in Figure~\ref{intlon} but it nevertheless traces the envelope of the proton intensities fairly well, except for a few large events from far behind the east limb, and events well behind the west limb that tend to lie below the curve. Thus, this may be a fair representation (other than being non-periodic) of the likely upper limit of the intensity of a 25 MeV proton event as a function of the longitude of the related solar event relative to the observer.  

\begin{figure} 
  \centering
   \includegraphics*[width=12cm,angle=0]{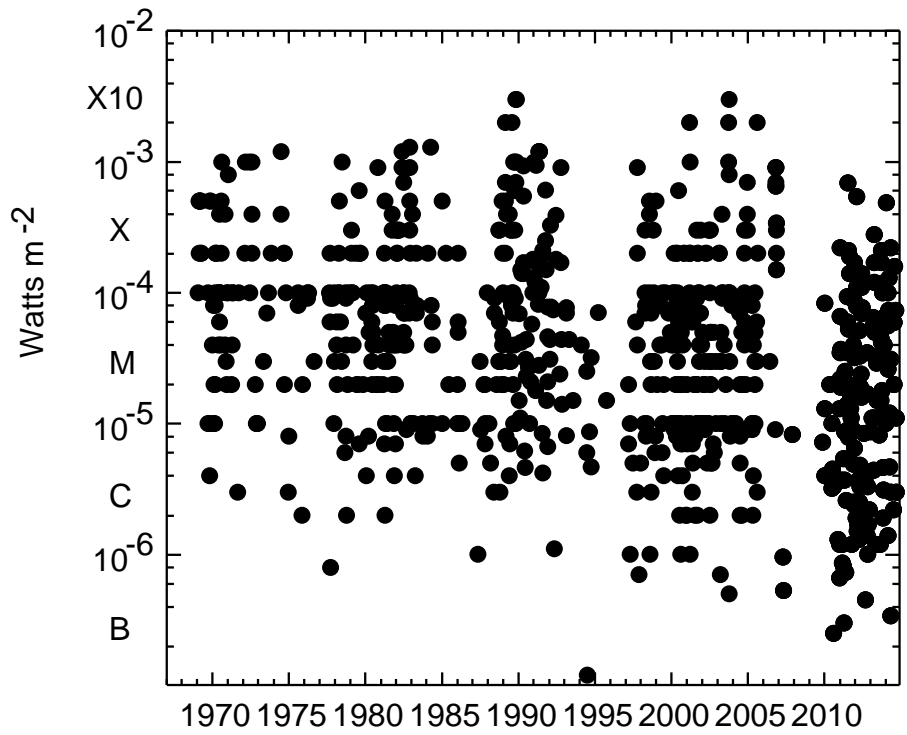}
  \caption{GOES soft X-ray peak intensity for 1054 flares associated with 25 MeV proton events in 1967--2014.}
  \label{xraytime}
\end{figure}

Currently, the only indicator of the solar event ``size" that we have for the complete period is the soft X-ray flare peak intensity, shown in Figure~\ref{xraytime} for 1054 flares associated with 25~MeV proton events.  (Note that for many events, the decimal fraction of the flare size is not currently recorded in our database, {\it cf.}, Table~1 of \cite{c10}.)  These flares range from class~B to class~X10, spanning around 4 orders of magnitude in peak soft X-ray intensity.  An intriguing feature is what appears to be a downward trend with time in the soft X-ray flare intensities associated with these proton events.  While there may be calibration effects in the soft X-ray data from the various GOES spacecraft that span this interval \citep{n11}, we also note that the number of proton events associated with small (B/low C-class) flares is apparently increasing with time. One possible reason is that it has become easier to make reliable associations with weak flares using the increasingly more comprehensive solar observations available during this period, including near-continuous solar and CME imaging from Cycle 23 and full-sun observations provided by the STEREOs in Cycle 24.  In particular, several 25~MeV proton events unambiguously associated with B-class flares were observed during Cycle 24 \citep{r14}.  When only limited solar observations are available, there is a tendency to choose a more intense flare from among several candidates, or to conclude that the related flare is uncertain or unknown, rather than select a weak flare.  Another contributor to the apparent downward trend is the lack of X10 class flares (so far) associated with proton events in Cycle 24, whereas such flares were evident in Cycles 22 and 23.  They were however, also rare in Cycle 21, and maybe in Cycle 20 though calibration may be an issue.  Nevertheless, even after allowing for these effects in the weakest and strongest flares, there is still an indication in Figure~\ref{xraytime} of an overall downward shift in the soft X-ray intensities of flares associated with 25~MeV proton events in Cycle 24 relative to previous cycles.

\begin{figure} 
  \centering
   \includegraphics*[width=10cm,angle=0]{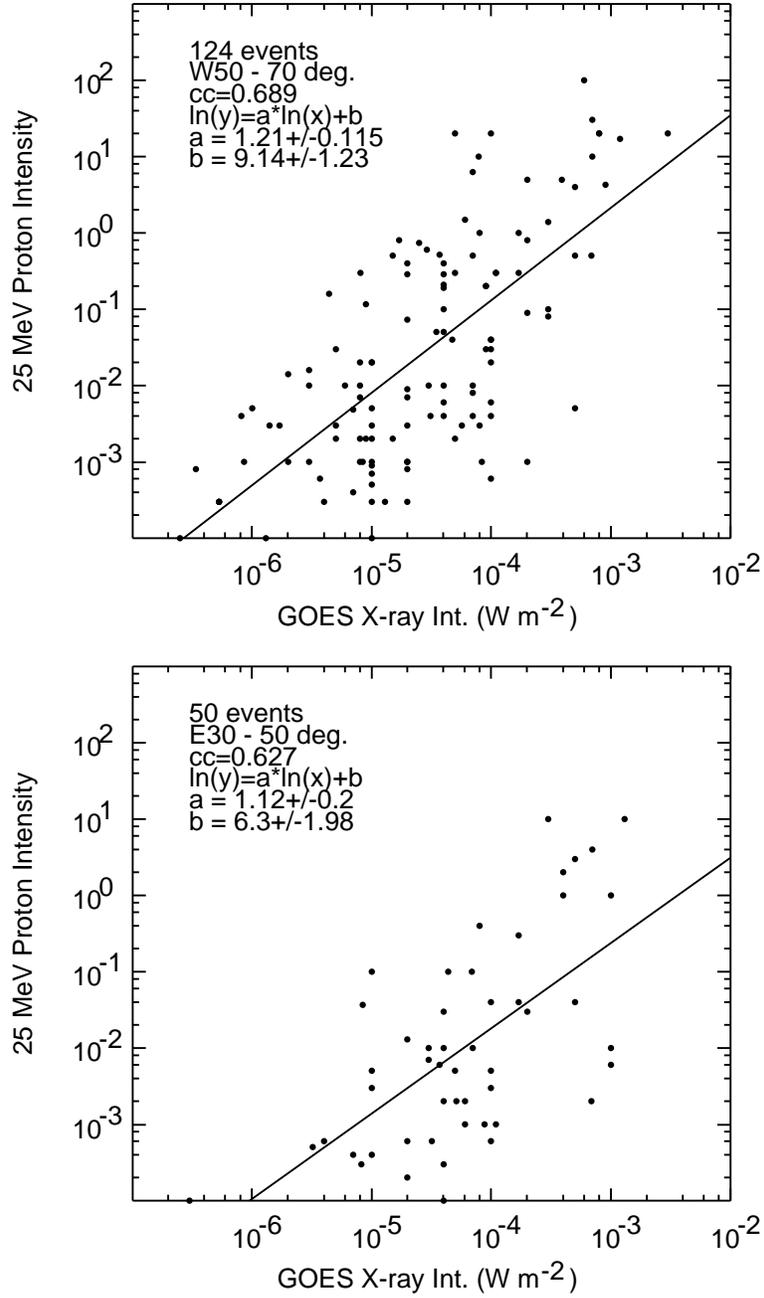}
  \caption{25 MeV proton event intensity ((MeV~s~cm$^2$~sr)$^{-1}$) vs. GOES soft X-ray peak intensity for 124 flares at W50--70$^\circ$ (top panel) and 50 flares at E30--50$^\circ$.}
  \label{xint}
\end{figure}

Figure~\ref{xint} illustrates that when a 25~MeV proton event is detected, the proton intensity is well correlated with the flare soft X-ray peak intensity (see also Figure~14 of \cite{r14}).  The top panel shows 124 well-connected events associated with flares at W50-$70^\circ$ longitude, with a correlation coefficient of 0.689.  With this large database of proton events, it is also possible to demonstrate such a correlation with a reasonable sample size even for poorly-connected events.  For example, the bottom panel of Figure~\ref{xint} shows a similar result for 50~events at E30--50$^\circ$ ($cc=0.627$); events in other longitude ranges also show similar correlations.  However, large X-ray flares are not necessarily accompanied by proton events at 1~AU.  For example, there were 164~M and X class flares on the western hemisphere in 2011--2013, and for at least 29 of these flares, the ongoing background particle intensities were low, yet no 25~MeV proton event was detected at Earth.  In addition, \cite{r16} discuss the low occurrence rate of 25 MeV proton events during the presence of the largest sunspot region in 24 years, in October 2014, that produced a number of confined X~class flares without coronal mass ejections \citep{s15, ch15}.

\section{Summary}
In this paper, we have summarized a few comparisons between SEP events that include 25 MeV protons in the current solar cycle (24) and previous cycles, using a large catalog of over a thousand such events detected by various spacecraft since 1967.  This  catalog is based on the event lists used in our previous studies, with events (e.g., weak or with unclear solar sources or data gaps) that were not included in these studies restored from scanning the original data sets.  Such a large data base allows SEP events during the current cycle to be placed in a wider context than just comparing those in Cycles~23 and 24.  It is planned to publish this catalog when all the relevant event information, such as the solar event identifications for restored events, is as complete as possible. The main points discussed here include:
\begin{itemize}

\item Cycle 24 is characterized by strong north-south hemisphere asymmetries in the temporal and spatial distributions of the solar events associated with 25~MeV proton events which follow those in the hemispheric sunspot numbers and areas (see also \cite{r16}).  In contrast, Cycle 23 was more symmetric between hemispheres.  In both cycles, there is a northern bias in the rising phase and early peak, changing to a southern bias, in the sunspot parameters and SEPs. 
\item The solar minimum between Cycles~23 and 24 was characterized by a lack of significant 25~MeV proton events, whereas such events were present during previous minima in the space era.
\item Proton events with intensities at 25 MeV similar to those of the most intense events detected at Earth in Cycles 23 and 22 were absent in Cycle 24, and also in Cycle 21, the largest cycle of the space era based on the smoothed sunspot number.
\item Comparing the large 25~MeV proton event and GLE rates in 1973--2007 suggests that around 5 to 7 $\pm1$ GLEs might have been expected in Cycle 24, to 2015, based on the proton event rate in this cycle, rather than the one, or possibly two, GLEs actually observed.   
\item Compared to the sunspot number, however, around twice as many 25~MeV proton events as might have been expected based on previous cycles have been detected in Cycle~24.
\item The envelope of the 25~MeV proton intensity-longitude distribution for over 1000 events is consistent with the average Gaussian fall off in intensity with longitude found in individual events observed at the STEREO spacecraft and at Earth, with a peak at well-connected longitudes.
\item A downward trend with time in the soft X-ray intensity of flares associated with 25~MeV proton events may be attributed in part to the improving ability to identify weaker flares associated with such events as solar observations have become more comprehensive together with the absence of the largest ($>$X10) SEP-associated flares in Cycle 24.
\item A positive correlation between flare soft X-ray and proton intensities is found for both well-connected and poorly connected events.         
\end{itemize}   

\section{Acknowledgments}
We thank the many researchers who have compiled the various data sets used in this paper.  The SOHO ERNE data are available from the Space Research Laboratory at the University of Turku (\url{http://www.srl.utu.fi/erne\_data/}).  The ACE data in Figure~\ref{event} are from the ACE Science Center (\url{http://www.srl.caltech.edu/ACE/ASC/}).  The STEREO High Energy Telescope data are available at \url{http://www.srl.caltech.edu/STEREO/Public/HET_public.html}.  The GLE list is maintained by the University of Oulu.  This work was supported by the NASA Heliophysics Living With a Star Science Program as part of the activities of the Focused Science Team ``Physics-based methods to predict connectivity of SEP sources to points in the inner heliosphere, tested by location, timing, and longitudinal separation of SEPs".









\end{document}